\documentclass[5p]{elsarticle}

\usepackage{comment}

\journal{Nuclear Instruments and Methods in Physics Research A}

\begin{document}

\begin{frontmatter}

\title{Lead fluoride Cherenkov detector read out by avalanche photodiodes for measuring the intensities of pulsed antiproton beams}

\author[UTokyo]{Y. Murakami\fnref{fn1}}
\fntext[fn1]{{\it Current address:} Hitachi Ltd. Corporation, Marunouchi 1-6-6, Chiyoda-ku, Tokyo, Japan.}
\author[MPQ]{H.~Aghai-Khozani\fnref{fn2}}
\fntext[fn2]{{\it Current address:} McKinsey and Company, Inc., Sophienstrasse 26, 80333 Munich, Germany.}
\author[MPQ]{M.~Hori\corref{mycorrespondingauthor}}
\cortext[mycorrespondingauthor]{Corresponding author}
\ead{Masaki.Hori@mpq.mpg.de}

\address[UTokyo]{Department of Physics, School of Science, University of Tokyo, 7-3-1 Hongo, Bunkyo-ku, 113-0033 Tokyo, Japan}
\address[MPQ]{Max-Planck-Institut fur Quantenoptik, Hans-Kopfermann-Strasse 1, 85748 Garching, Germany}

\begin{abstract}
A Cherenkov detector based on an array of five lead fluoride ($\beta$-PbF$_2$) crystals of size 30 mm$\times$30 mm$\times$160 mm
read out by reverse-type avalanche photodiodes (APD's) of active area 10 mm$\times$10 mm was used to measure the 
flux of secondary particles emerging from the annihilation of pulsed beams of antiprotons at the Antiproton 
Decelerator of CERN. We compared the relative photon yields of radiators made of $\beta$-PbF$_2$, fused silica, 
UV-transparent acrylic, lead glass, and a lead-free, high-refractive-index glass. 
Some {\it p-i-n} photodiodes were also used for the readout, but the output signals were dominated by the
nuclear counter effect (NCE) of secondary particles traversing the 300 $\mu{\rm m}$ thick
depletion regions of the photodiodes. Smaller NCE were observed with the APD's, as the maximum electronic gain in them
occurred predominately for electron-ion pairs that were generated in the thin ${\it p}$-type semiconductor layer that proceeded the
{\it p-n} junction of high electric field where amplification took place.
\end{abstract}


\end{frontmatter}

\section{Introduction}
\label{sec:introduction}

The Atomic Spectroscopy and Collisions Using Slow Antiprotons (ASACUSA) collaboration has recently measured the 
cross sections $\sigma^A_{\rm tot}$ of antiprotons with an incident kinetic energy $E\sim 5.3$ MeV annihilating via
in-flight nuclear reactions in carbon, Mylar, nickel, tin, and platinum targets \cite{bianconi2011,corradini2013,aghai2018}. 
Measurements of $\sigma^A_{\rm tot}$ were also attempted on $E=130\pm 10$ keV antiprotons annihilating in 
carbon, palladium, and platinum target foils of sub-100 nm thicknesses \cite{aghai2012,soter2014,todoroki2016}. These 
experiments were carried out at the Antiproton Decelerator (AD) of CERN using $\Delta t=50$--120 ns long pulsed beams 
that contained between $N_{\overline{p}}=2\times 10^5$ and $5\times 10^6$ antiprotons. 
The arrival of individual 
antiprotons at the target could not be resolved. We instead determined the relative intensity and temporal structure of the beam by using 
acrylic Cherenkov detectors to measure the flux of secondary charged particles (mostly charged pions, muons, and electrons)
that emerged from the simultaneous annihilations in the apparatus. These past experiments utilized
fine-mesh photomultiplier tubes \cite{mhori2003-2} which had signal responses that were highly linear to
read out the Cherenkov radiators.

\begin{table*}[tbp]
\caption{Specifications of the avalanche and {\it p-i-n} photodiodes used in this study:
their active areas, effective thicknesses against ionizing radiation, terminal capacitances, 
cutoff frequencies of the output signals, and wavelength ranges over which the photodiodes retain sensitivity.}
\begin{center}
  \begin{tabular}{cccccc} 
     \hline\hline
     Photodiode & Area & Eff. thickness & Capacitance  & Cutoff freq. & Spect. response\\ 
                         & (mm$\times$mm)    & ($\mu$m) & (pF) & (MHz) &  (nm) \\ \hline
     S8664-1010 (APD) & 10$\times$10 & $\sim 5$ & 270  & $\sim$11 & 320--1000 \\
     S3204-08 ({\it p-i-n}) &  18$\times$18 & 300   & 130 & 20 & 340--1100 \\
     S3590-08 ({\it p-i-n}) &  10$\times$10 & 300   & 40   & 40 & 340--1100 \\ \hline\hline
       \end{tabular}
       \end{center}
\label{table:types}
\end{table*}

In this paper, we describe a different Cherenkov detector that consisted of an array of five lead fluoride crystals 
in the cubic phase ($\beta$-PbF$_2$) \cite{Anderson1990385,Woody256613,Appuhn1994208,Achenbach2001318} 
of size 30 mm$\times$30 mm$\times$160 mm. The arrival of antiproton pulses 
at the experimental target produced an intense flash of Cherenkov light in each crystal; this was measured by a reverse-type \cite{kirn1997,lecomte1999,cms2000,kataoka2005,ikagawa2005} 
avalanche photodiode (APD) of active area 10 mm $\times$10 mm. These crystals are sensitive to both the charged secondary particles
and $\gamma$-rays \cite{mhori2003-2} that emerge from antiproton annihilations, and are expected to have a higher 
light yield compared to Cherenkov detectors based on acrylic radiators of the same size. Other advantages of the detector
include its compact size and insensitivity to magnetic fields, which may allow its use in experiments
involving magnetic Penning traps within the physics program of the AD \cite{mhori2013}. 
The radiation-hard nature of the $\beta$-PbF$_2$ crystals may allow their use in measuring 
high-intensity beams of electrons in laser spectroscopy experiments of metastable pionic 
helium atoms \cite{mhori2014}. We compared the light yields of Cherenkov radiators made of $\beta$-PbF$_2$, 
fused silica, acrylic, lead glass, and a lead-free, high-refractive index glass against antiproton annihilations. 
We also attempted to use {\it p-i-n} photodiodes \cite{hosono1995} for the readout, but the waveform of the
output signals contained a dominant background that arose from the nuclear counter effect (NCE) of secondary 
particles traversing the $\sim 300$ $\mu{\rm m}$ thick depletion regions of these detectors \cite{Satpathy1997423,mao2011}. 
The relative size of the NCE was much smaller in reverse-type APD's, because the maximum electronic gain occurred predominantly 
for those electron-ion pairs that were generated in the thin ${\it p}$-type semiconductor layer that proceeded the
{\it p-n} junction of high electric field where amplification took place \cite{kirn1997,lecomte1999,cms2000,kataoka2005,ikagawa2005,mao2011}. 
The detector was recently used in an experiment to determine the cross section 
$\sigma^A_{\rm tot}$ of 5.3 MeV antiprotons annihilating in a carbon target \cite{aghai2018}.

\begin{figure}[btp]
\begin{center}
\includegraphics[width=80mm]{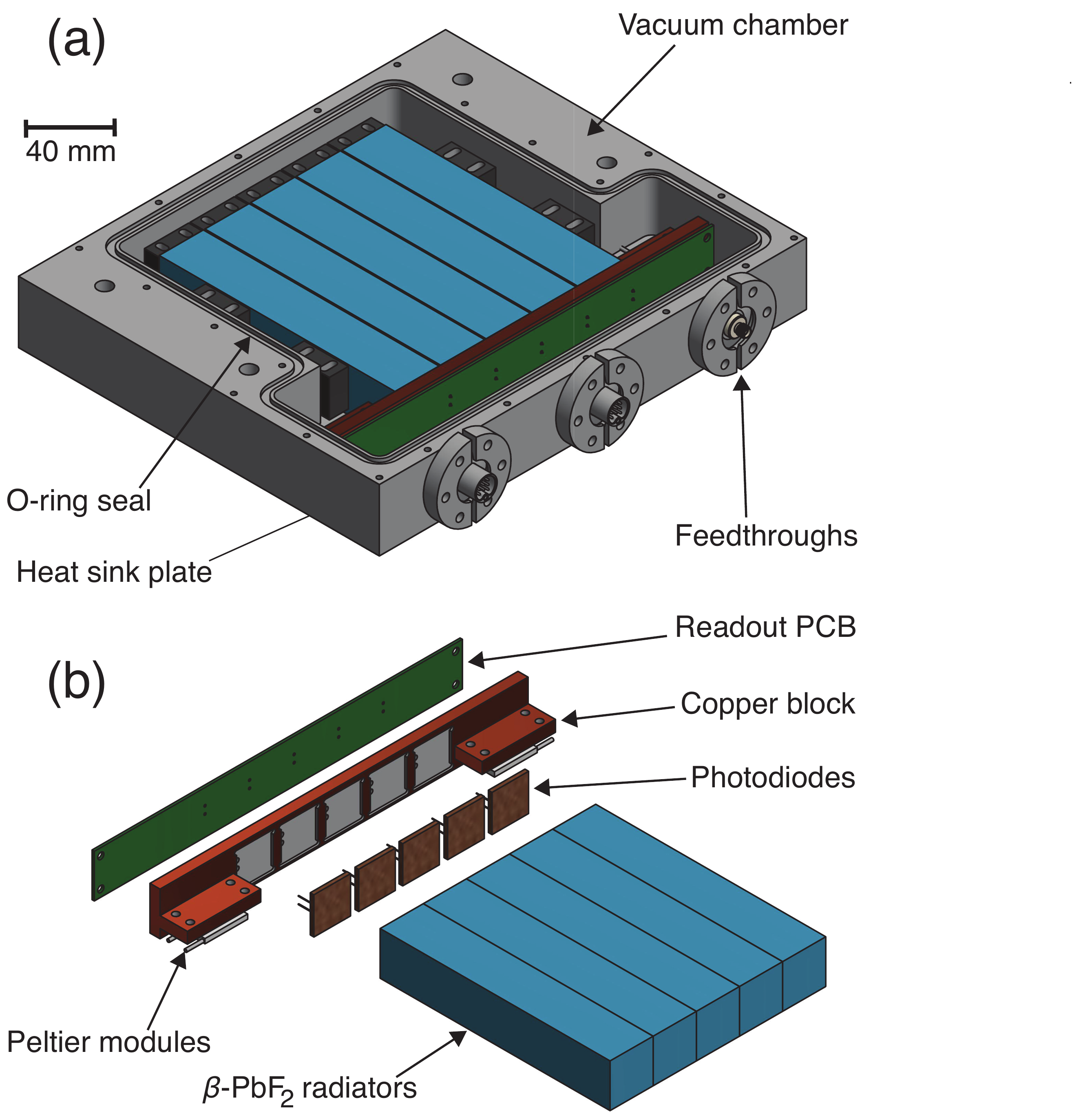}
\end{center}
\caption{(a) Layout of the Cherenkov detector hermetically sealed in an aluminum chamber. (b) Details of the $\beta$-PbF$_2$ radiator, photodiodes, copper block, printed circuit board (PCB), and Peltier thermoelectric modules.}
\label{fig:detector}
\end{figure}

Since the 2000's, $\beta$-PbF$_2$ crystals have been used in electromagnetic calorimeters with high timing 
resolutions. The A4 collaboration used 1022 crystals of lengths between $l=150$ and 180 mm, in a 
parity-violating, electron-scattering experiment that was carried out at the Mainz Microtron (MAMI) facility 
\cite{Achenbach2001318,baunack2009}. Each crystal was read out by a photomultiplier tube with 10-stage, 
line focusing dynodes. The muon g-2 experiment at Fermilab utilized $\sim 1300$ crystals of size 
25 mm$\times$25 mm$\times$140 mm,
which were read out by arrays of 16 silicon photomultipliers (SiPM's) of active area 12 mm$\times$12 mm
\cite{Fienberg201512,Kaspar2017}. Positrons of energy $E\sim 2$ GeV that emerge from $\mu^+$ decays in a storage ring 
were reconstructed with an energy resolution of $<5\%$ and a timing resolution $\Delta t<100$ ps.
A positron annihilation lifetime spectroscopy (PALS) experiment used 
the $\beta$-PbF$_2$ crystals to detect 511 keV gamma rays with a timing resolution $\Delta t\sim 3.5$ ns
\cite{Cassidy20071338}. 

The advantages of $\beta$-PbF$_2$ crystals include a high refractive index ($n_r=1.82$) and density 
($\rho=7.77$ g/cm$^3$), significant transmission at ultraviolet wavelengths $\lambda\sim 260$ nm, radiation hardness ($\sim 10^4$--$10^5$ rad) \cite{Anderson1990385,kobayashi2001}, and a small radiation length ($\ell_r=9.3$ mm) and Moliere radius ($M_r=22$ mm). 
On the other hand, the light yield of $\beta$-PbF$_2$ is much smaller than those of scintillators such as 
lead tungstate (PbWO$_4$) and thallium activated cesium iodide [CsI(Tl)]. For each MeV of energy deposited in
a $\beta$-PbF$_2$ crystal, a photomultiplier coupled to it typically detects $\Gamma=1.5$--1.9 photoelectrons \cite{Achenbach2001318}. 
It is difficult to grow large ($\ell>>200$ mm) crystals of high transparency and optical uniformity
\cite{Achenbach907578,Ren2002539,Ren2003141}. It was noted that when oxygen impurities were allowed to
diffused into the crystals during their growth, regions that consist of lead oxide, lead fluoro-oxide, or the opaque, 
orthorhombic phase of lead fluoride ($\alpha$-PbF$_2$) were formed. These appeared as scattering centers and inclusions 
that could reduce the collection efficiency of the Cherenkov light. It has also been reported that when
exposed to a high-humidity environment, the surfaces of the crystals changed into a frosty-whitish appearance \cite{Ren2002539,Ren2003141}. 

\begin{figure}[tbp]
 \begin{center}
 \includegraphics[width=80mm]{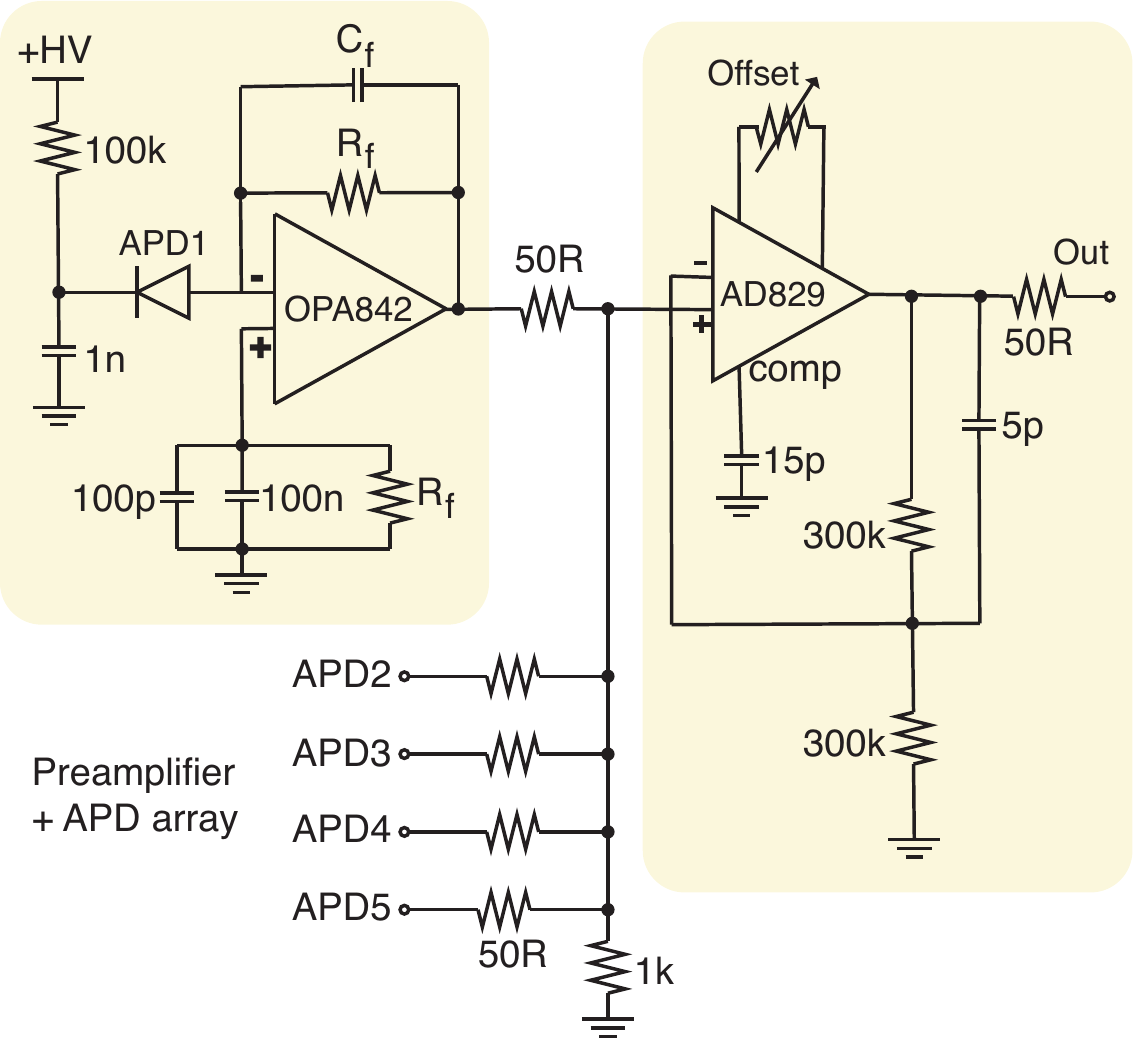}
 \end{center}
 \caption{Circuit diagram of the array of five transimpedance preamplifiers of which one channel is shown. 
 The output signals are summed using a resistor network and a driver amplifier. The feedback resistor $R_f$ and capacitor $C_f$ of 
 each preamplifier were adjusted to optimize the timing response and stability of the APD or {\it p-i-n} photodiode signal.}
\label{fig:circuit_cherenkov}
\end{figure}

\section{Detector}

\subsection{Construction}

The five $\beta$-PbF$_2$ crystals of size 30 mm$\times$30 mm$\times$160 mm (see Fig.~\ref{fig:detector}) 
were wrapped in two layers of a polyvinylidene fluoride membrane with 0.45 $\mu{\rm m}$ pores (Millipore Immobilone-P), 
which constituted a diffusive reflector. A silicon APD of active area 10 mm$\times$10 mm (Hamamatsu Photonics S8664-1010, see Table~\ref{table:types}) was attached to one end of each crystal. The five APD's in 
the array were thermally anchored on a 200 mm long copper block using thermal paste 
(Electrolube HTCP). A two-terminal, temperature-to-voltage transducer (Analog Devices AD590MH) was also glued to the block using 
an epoxy adhesive that had a high thermal conductivity 
(Henkel Loctite Stycast 1266). This assembly was cooled to a temperature $T\sim 19$ $^{\circ}$C with a stability of 
$\pm 0.1$$^{\circ}$C, using two Peltier thermoelectric modules (Adaptive ET-03-10-13-RS) of size 
30 mm$\times$15 mm and cooling power $\sim 6$ W that were regulated by a proportional--integral--differential controller
(Thorlabs TED200C) and a separate 5 W direct current (DC) power supply. 
The electronics for biasing the photodiodes, amplifying the output signals, and providing the low-noise, regulated DC 
power sources of $\pm 5$ V and $\pm 15$ V were integrated on a four-layered printed circuit board (PCB) of size 
30 mm$\times$200 mm. This PCB made of 1.6 mm thick glass epoxy (Panasonic Electronic Works R1705) 
was also mounted to the copper block using four Teflon standoffs that helped to thermally isolate it from the block.

The detector was housed in a 290 mm$\times$245 mm$\times$50 mm aluminum chamber of wall thickness 
$t_r=5$ mm. The chamber was hermetically sealed and purged with nitrogen to prevent condensation 
on the cooled photodiodes and $\beta$-PbF$_2$ crystals that may arise from the high humidity in the vicinity of the cryogenic target. 
The chamber also shielded the photodiodes against the electromagnetic interference of frequency range $f=10$--200 MHz 
that was emitted from the AD and radiofrequency quadrupole decelerator (RFQD) \cite{mhori2013}.
Detector modules which employed silicon {\it p-i-n} photodiodes of active area 18 mm$\times$18 mm (S3204-08) or 
10 mm$\times$10 mm (S3590-08) instead of the APD were also constructed (Table~\ref{table:types}).
The photodiodes were protected by thin windows of epoxy resin. 

A bias voltage $V_{\rm b}\sim 380$ V was applied to the cathode terminal of each APD through a 
low-pass filter that consisted of a $R=100$ k$\Omega$ thin metal film resistor and a $C=1$ nF 
radiofrequency mica capacitor (see Fig.~\ref{fig:circuit_cherenkov}).
The anode terminals were connected to transimpedance preamplifiers that contained an operational amplifier 
(Texas Instruments OPA842IDB) with a gain bandwidth product of $f_b=200$ MHz, 
input voltage noise 2.6 nV/$\sqrt{\rm Hz}$, low distortion, and a DC voltage offset $\pm 0.3$ mV.
A feedback resistor $R_f=$1 k$\Omega$ which corresponded to a photocurrent-to-voltage conversion
ratio $g\sim 10^3$ V/A  was chosen to provide the best dynamic range for measuring the Cherenkov signals
with typical amplitudes of 20--40 mV.
As the APD's had a large terminal capacitance $C_d\sim 270$ pF, a compensating feedback capacitor $C_f=22$ pF
was added to avoid excessive ringing and oscillation of the output signal, and obtain
a flat frequency response with a -3dB corner frequency $f_r\sim\sqrt{f_b/2\pi R_fC_d}\sim 6$ MHz.
For reading out the S3204-08 {\it p-i-n} photodiodes of $C_d\sim 130$ pF, the feedback elements were changed to values
$R_f=3$ k$\Omega$ and $C_f=12$ pF which provided the best compromise between a higher gain 
$g\sim 3\times 10^3$ V/A, corner frequency $f_r\sim 9$ MHz, and the stability of the transimpedance preamplifier. For the smaller 
{\it p-i-n} photodiodes (S3590-08) of $C_d\sim 40$ pF, an operational amplifier (OPA843IDB) of larger gain 
bandwidth product $f_b=800$ MHz in conjunction with feedback values $R_f=3$ k$\Omega$ and $C_f=$ 2 pF 
were used to obtain a -3dB corner frequency $f_r\sim 20$ MHz.

The output signals of the five transimpedance preamplifiers were summed 
using a resistor network, and amplified by a non-inverting follower of gain 2 which was based on a bipolar operational amplifier 
(Analog Devices AD829AR) with an input voltage noise $\sim 1.7$ nV/$\sqrt{\rm Hz}$. This resulted in an output signal
of typical amplitude $V=200$--400 mV into a 50 $\Omega$ impedance, which was superimposed on a DC 
offset. The signal was transmitted out of the chamber using a SMA-type coaxial vacuum feedthrough, and its analog waveform 
recorded by a digital oscilloscope of analog bandwidth 
$f=4$ GHz, digital sampling rate $f_s=10$ Gs/s, and a vertical resolution of 8 bits (Lecroy Waverunner 640Zi). 
The $\pm 5$ V and $\pm 15$ V DC power
needed for the amplifiers were generated by four linear regulators (Linear Technology LT1965IDD and LT3015IDD) and 
passed through some electromagnetic interference filters (Murata BNX002). The linear regulators were supplied with 
source voltages that were adjusted slightly ($\sim 0.35$ V) above their regulated output voltages. This helped to minimize 
the heat dissipation in the chamber.

\subsection{Cherenkov radiators}

\begin{table*}[tbp]
\caption{Refractive indices, radiation lengths, and measured ultraviolet (UV) cutoff wavelengths of
five types of Cherenkov radiator materials used in this work. The UV cutoff wavelengths are approximate
values based on the measurements shown in Fig.~\ref{fig:transmission_data}.}
\begin{center}
  \begin{tabular}{ccccc} \hline\hline
      Radiator & Refractive index & Density                  & Radiation length & UV cutoff \\ 
                    &                            &  (g/cm$^3$)           &  (mm)                 &   (nm)                 \\ \hline
      $\beta$-PbF$_2$ (lead fluoride)                                   & 1.82 & 7.77 & 9.3             &  $\sim 260$\\
      T-4040 (fused silica)                             & 1.49 & 2.2 & $\sim 120$  & $<190$ \\ 
      Clarex S-0 (acrylic)                             & 1.46 & 1.18 & $\sim 340$ & $\sim 320$ \\
      SF57HTultra (lead glass)                     & 1.85 & 5.51 & 15               &    $\sim 390$ \\
      S-TIH53W                                              & 1.84 & 3.54 &                    &  $\sim 400$\\ \hline\hline
       \end{tabular}
\end{center}
\label{table:radiators}
\end{table*}

\begin{figure}[tbp]
\centering
\includegraphics[width=75mm]{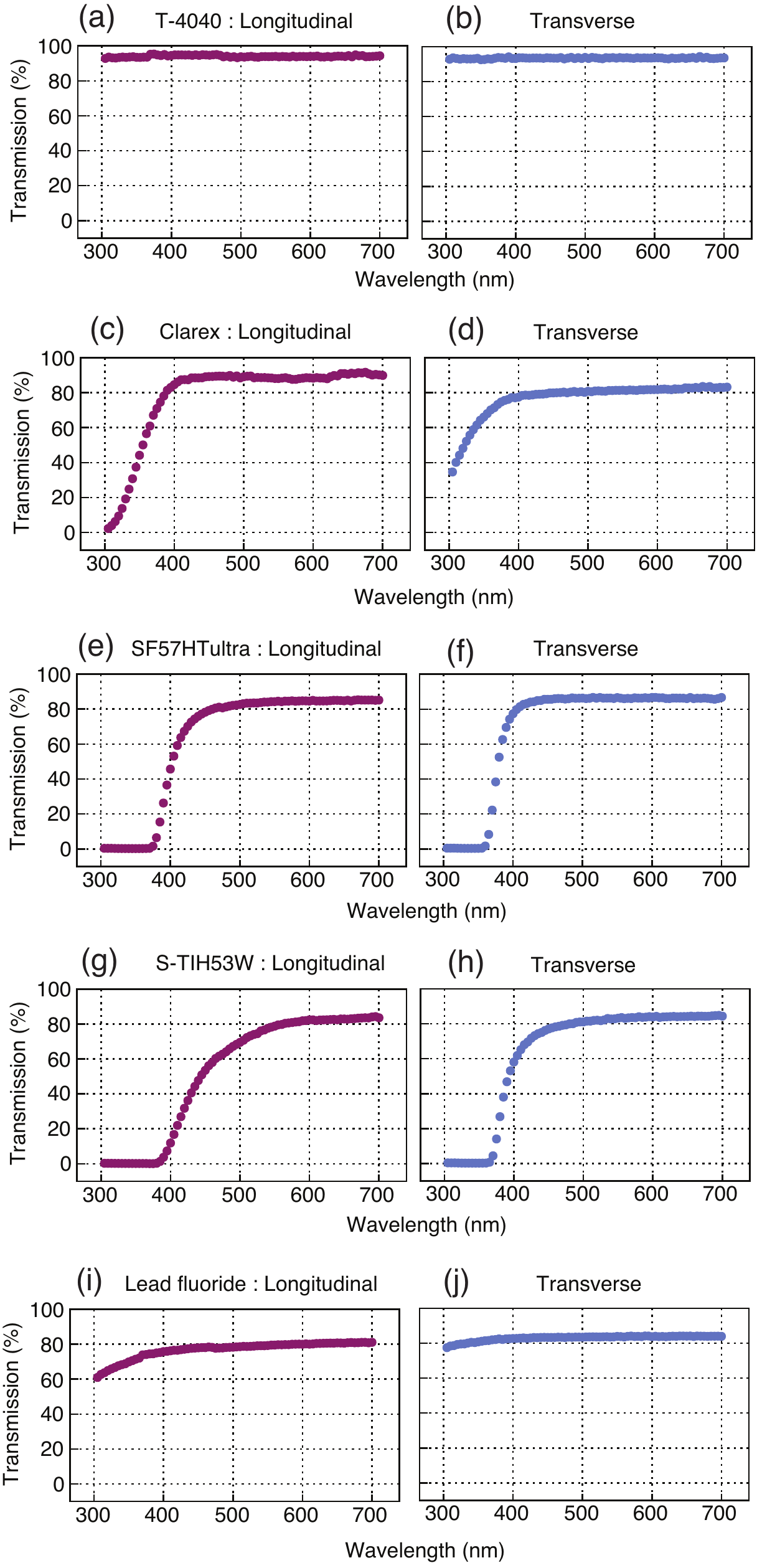}
\caption{Longitudinal and transverse transmissions of Cherenkov radiators made of fused silica (T-4040), acrylic (Clarex S-0), 
lead glass (SF57HTultra), and lead free high refractive index glass (S-TIH53W) at wavelengths between $\lambda=300$ nm and 700 nm measured with a photospectrometer (see text).}
\label{fig:transmission_data}
\end{figure}

The $\beta$-PbF$_2$ crystals were grown using the Bridgman method and all surfaces polished 
by the Shanghai Institute of Ceramics, Chinese Academy of Sciences (SICCAS). The crystals were specified by
the manufacturer to be without cracks or scratches longer than 10 mm and deeper than 2 mm, and no inclusions, 
scattering centers, or microcracks observable with the naked eye. Cherenkov radiators of the same size made of 
fused silica (CoorsTek T-4040), UV transparent grade acrylic plastic (Nitto Jushi Kogyo Co. Clarex S-0),
high transmission grade lead glass (Schott AG SF57HTultra), and Ohara S-TIH53W glass were
also manufactured (see Table. \ref{table:radiators}). The last of these was a lead-free glass with a high refractive
index $n_r\sim 1.8$, density $\rho=3.5$ g/cm$^3$, and partial dispersion of Abbe number $\nu_d\sim 23.8$. 
X-ray photoelectron spectroscopy measurements on ground samples of the glass showed that it contained 
barium, titanium, and niobium oxides. These ``eco-friendly" alternatives to conventional flint glass 
generally have reduced transmission at blue and UV wavelengths. The surfaces of the fused silica, 
lead glass, and S-TIH53W glass radiators were optically polished to a quality that was specified to be
scratch-dig 60-40 and flatness of peak-to-valley $\sim \lambda/2$ wave.

\begin{figure}[tbp]
\centering
\includegraphics[width=70mm]{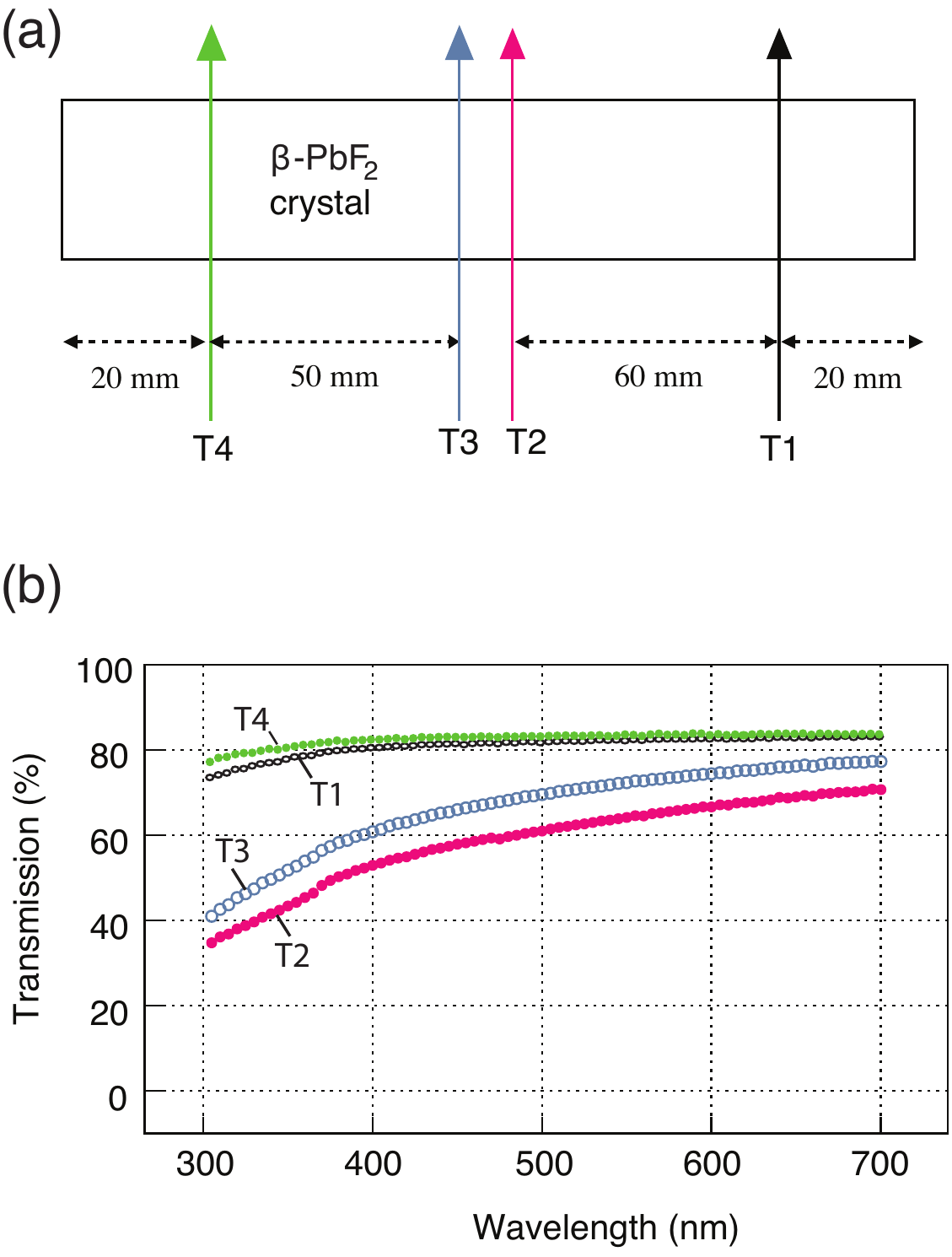}
\caption{(a) Four paths (indicated by T1--T4) parallel to the transverse 
axis of the $\beta$-PbF$_2$ crystals where the optical transmissions were measured. 
(b) Transmissions at wavelengths between $\lambda=300$ and 700 nm corresponding to each light path.}
\label{fig:transmissionpbf2}
\end{figure}

The optical transmissions (Fig.~\ref{fig:transmission_data}) of each radiator along its longitudinal (i.e., long) axis, and at four 
positions [indicated by arrows labeled T1, T2, T3, and T4 in Fig.~\ref{fig:transmissionpbf2}(a)] parallel to its 
transverse (short) axis at wavelengths between $\lambda=300$ and 700 nm were measured using a photospectrometer. 
The device was of single beam type and included a xenon halide lamp as the light source.
No position-dependent variations over T1--T4 were seen in the transverse transmissions of the fused silica, 
acrylic, lead glass, and S-TIH53W radiators, and so we plotted the typical values for each material in
Figs.~\ref{fig:transmission_data}(b), \ref{fig:transmission_data}(d), \ref{fig:transmission_data}(f), and \ref{fig:transmission_data}(h), respectively.

The fused silica radiators [Figs.~\ref{fig:transmission_data}(a)--(b)] showed high transverse and longitudinal transmissions 
of $T_t\sim T_l\sim 94\%$ over the entire range of measured wavelengths. The majority of the losses were caused by 
the $\sim 3\%$ reflections at each glass-to-air interface, whereas the attenuation in the glass was negligible. 
The transverse and longitudinal transmissions of the acrylic radiator at visible wavelengths 
$\lambda=400$--700 nm were slightly lower at $T_t\sim 90\%$ and $T_l\sim 80\%$, respectively
[Figs.~\ref{fig:transmission_data}(c)--(d)]. Significant reductions of 
the transmissions to $T_t\sim 40\%$ and $T_l\sim 15\%$ were seen at a UV wavelength $\lambda\sim 330$ nm. 

The lead glass radiators [Fig.~\ref{fig:transmission_data}(e)] retained transmissions of $T_l\sim 82\%$ over 
the visible wavelengths $\lambda=490-700$ nm, which rapidly decreased to $\sim 40\%$ and $\sim 10\%$ 
at the UV wavelengths $\lambda=400$ nm and $380$ nm, respectively.
The transverse transmission [Fig.~\ref{fig:transmission_data}(f)] was somewhat higher: 
$T_t\sim 83\%$, $\sim 40\%$, and $\sim 10\%$ at the respective wavelengths $\lambda=420-700$ nm, 
$380$ nm, and $370$ nm. The S-TIH53W radiator had a slightly yellowish color [Figs.~\ref{fig:transmission_data} (g)--(h)],
and the highest absorption among the materials studied here. The longitudinal transmissions at wavelengths
$\lambda=590-700$ nm and 405 nm were $T_l\sim 80\%$ and $\sim 20\%$, respectively. The transverse values 
at $\lambda=520-700$ nm and 380 nm were $T_t\sim 82\%$ and $\sim 20\%$, respectively.

The UV transmission of the $\beta$-PbF$_2$ crystals were lower compared to fused silica, but higher 
than the other three materials studied here [Figs.~\ref{fig:transmission_data}(i)--(j)]. 
Small, white scattering centres along the surfaces of the $\beta$-PbF$_2$ crystals caused significant 
variations in the optical transmission: in one sample, transmissions 
of $T_t\sim$74\%, $\sim$35\%, $\sim$32\%, and $\sim$78\% were measured at the respective positions 
T1, T2, T3, and T4 [Fig.~\ref{fig:transmissionpbf2}(b)]. The longitudinal transmission was typically 
$T_l\sim$ 80\% at visible wavelengths $\lambda=400$--700 nm, and $T_l\sim 60\%$ at $\lambda=300$ nm. 

\subsection{APD's and {\it p-i-n} photodiodes}

The S8664-1010 silicon APD's \cite{kataoka2005,ikagawa2005} (Table~\ref{table:types}) of active area 10 mm $\times$ 10 mm
were previously used to read out CsI(Tl), BGO, and GSO(Ce) scintillators that emit at blue and green wavelengths 
\cite{cms2000,jin2016,comet}. The quantum efficiencies specified at wavelengths 
$\lambda=330$ nm, 420 nm, and 600 nm were $\varepsilon\sim 45\%$, $70\%$, and $85\%$, respectively. 
The APD's were of the so-called reverse type, where the {\it p-n} junction of high field in which multiplication
takes place was typically located about $\sim 5$ $\mu{\rm m}$ from the illuminated surface of the sensor
\cite{kirn1997,lecomte1999,cms2000,kataoka2005,ikagawa2005}.
Visible photons were absorbed and generated electron-hole pairs within the first few microns of the depletion region. 
These electrons were collected and underwent multiplication with the highest gain. Although charged particles, X-rays, and neutrons 
traversing the APD generated primary ion pairs over the entire width of the photodiode, 
only those pairs that originated in the thin {\it p}-type layer preceding the high-field {\it p-n} junction underwent high-gain multiplication.
This lead to a small ($t_r<6$ $\mu$m) effective thickness of the device with regards to the nuclear counting effect (NCE) 
against ionizing radiation, which was measured using a $^{90}$Sr source \cite{cms2000}.

The S3204-08 silicon {\it p-i-n} photodiode, on the other hand, had an active area of 18 mm $\times$ 18 mm,
depletion layer thickness $t_d=300$ $\mu$m, terminal capacitance $C_d\sim 130$ pF, and sensitivity at 
wavelengths between $\lambda=340$ and 1100 nm with a maximum at $\lambda\sim 960$ nm (Table~\ref{table:types}). 
Similar devices have been previously used for CsI(Tl) calorimeter readout \cite{ikeda2000,imazato2000,aubert2002,ablikim2010}.
The S3509-08 {\it p-i-n} photodiode had a smaller active area $10$ mm$\times$10 mm and capacitance
$C_d\sim 40$ pF, and the highest signal cutoff frequency $f_c=40$ MHz among the detectors used here.

\begin{figure}[tbp]
\centering
\includegraphics[width=80mm]{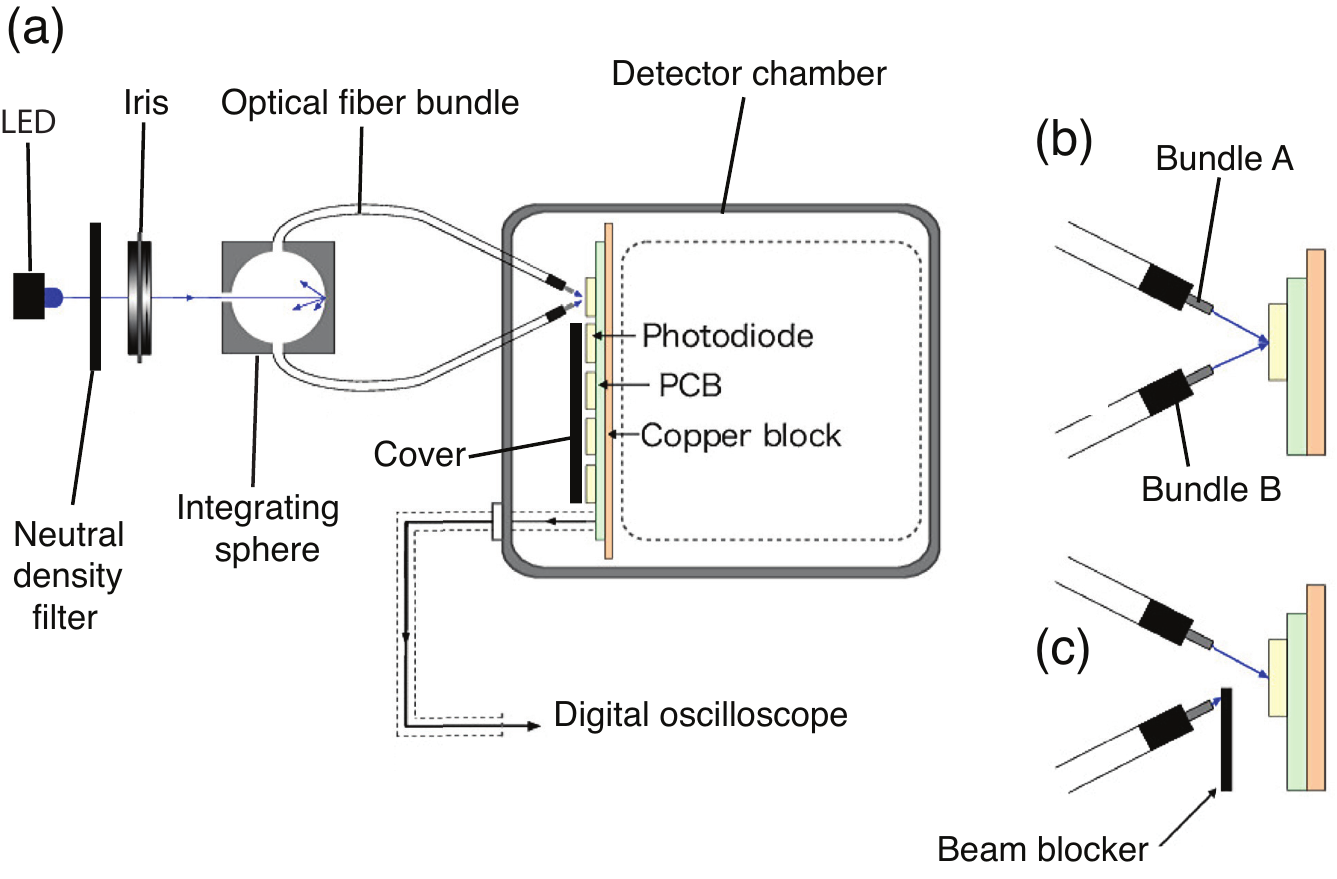}
\caption{(a) Layout for measuring the linearity response of the photodiode and transimpedance preamplifier combinations, not to scale. Blue light pulses produced by a LED were attenuated by neutral density filters before entering an integrating sphere. The light was extracted from the sphere through a pair of optical fiber bundles. (b) The light emitted from the two fiber bundles A and B were focused onto the surface of one of the photodiodes in the array. The remaining four photodiodes were masked using a sheet of black rubber (see text). (c) For some of the measurements, a sheet of black rubber was used to prevent the light emitted by fiber B from reaching the photodiode.}
\label{fig:linearity_test}
\end{figure}

\begin{figure}[tbp]
\centering
\includegraphics[width=80mm]{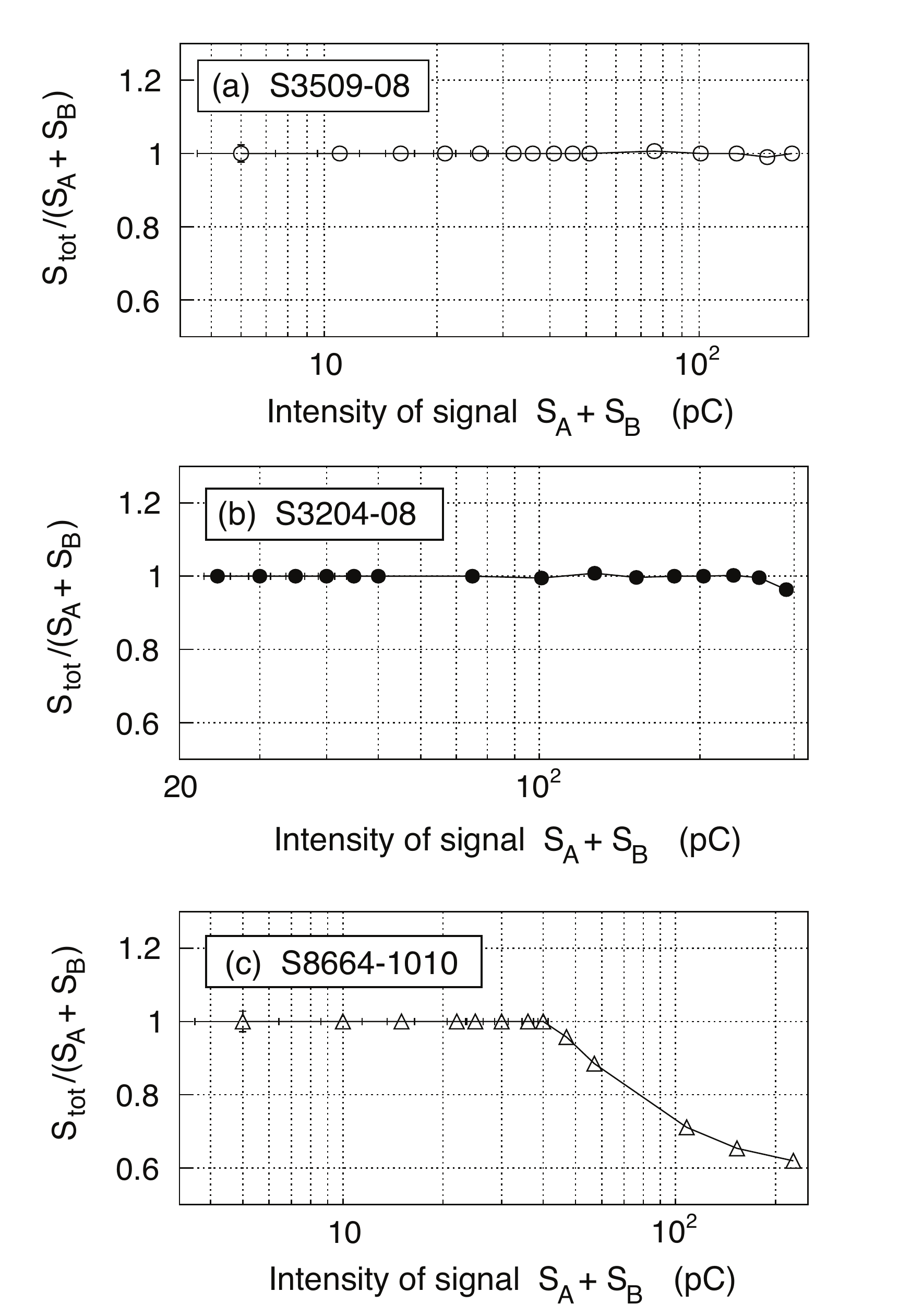}
\caption[linearity]{Linearity responses of the {\it p-i-n} photodiodes (a) S3590-08 and (b) S3590-08, and (c) the APD S8664-1010 combined with the transimpedance preamplifiers shown schematically in Fig.~\ref{fig:circuit_cherenkov} measured using a LED light source, see text.}
\label{fig:linearity_test_result}
\end{figure}

The signal linearity responses of the three photodiode and transimpedance preamplifier combinations were 
measured \cite{Budde:79, Mielenz:72} using the setup of Fig.~\ref{fig:linearity_test}(a). A light emitting diode 
(Thorlabs M450D3) produced a 100 ns long light pulse of emission wavelength $\lambda\sim 450$ nm, 
power $\sim 300$ mW, and emission angle $\theta=120^{\circ}$. The light was attenuated using 
a pair of variable neutral density filter wheels and an iris, before being coupled into a 51 mm diameter 
integrating sphere (Thorlabs IS200) that had an internal surface reflectivity of 99\%. The propagation direction
and polarization of the light was thus randomized. Part of this light was extracted from the sphere
through a pair of optical fiber bundles (Thorlabs BF20HSMA01) that each consisted of 7 multimode optical fibers of core 
diameter $d\sim 550$ $\mu$m and numerical aperture $N_A=0.22$. The fibers were made of fused silica with a 
high concentration of hydroxyl groups (OH). The intensity of the light emitted from each fiber bundle [indicated by A and B in
Fig.~\ref{fig:linearity_test}(b)] was found to be equal to within $\sim 10\%$. The light was 
expanded using aspheric lenses of focal length $f_l\sim 18$ mm (Thorlabs F280SMA-A) into a 
8 mm diameter spot on the surface of one of the photodiodes. The other four photodiodes in the array 
were masked with a sheet of black rubber. The APD's were biased at the nominal
voltage $V_b=380$ V. Measurements were carried out by irradiating the photodiode with, 
a): only the light beam emitted from bundle A [see Fig.~\ref{fig:linearity_test}(c)] corresponding to a measured 
signal intensity $S_A$ at the output of the readout circuit of Fig.~\ref{fig:circuit_cherenkov}, b): the light
from bundle B corresponding to an intensity $S_B$, and c): simultaneous irradiation by the two beams corresponding to
a measured signal intensity $S_{\rm tot}$ [Fig.~\ref{fig:linearity_test}(b)]. 
Any saturation in the photodiode or preamplifier would lead to the $S_{\rm tot}$ value 
being smaller than the expected sum $S_A+S_B$. In Fig.~\ref{fig:linearity_test_result} the 
measured fraction $S_{\rm tot}/(S_A+S_B)$ at various sums $S_A+S_B$ of the driver amplifier output 
(Fig.~\ref{fig:circuit_cherenkov}) in pC are shown for each photodiode type. Both {\it p-i-n} photodiode-amplifier 
combinations remained linear up to $S_A+S_B\sim 200$ pC. Any residual nonlinearity
is primarily caused by the transimpedance and buffer amplifiers. The APD's on the other hand 
saturated at a signal output of $>40$ pC, presumably due to space-charge effects in the depletion region of the
APD reducing the effective gain. 

\section{Measurements}

\subsection{Relative light yields of Cherenkov radiators}

\begin{figure}[tbp]
 \begin{center}
 \includegraphics[width=85mm]{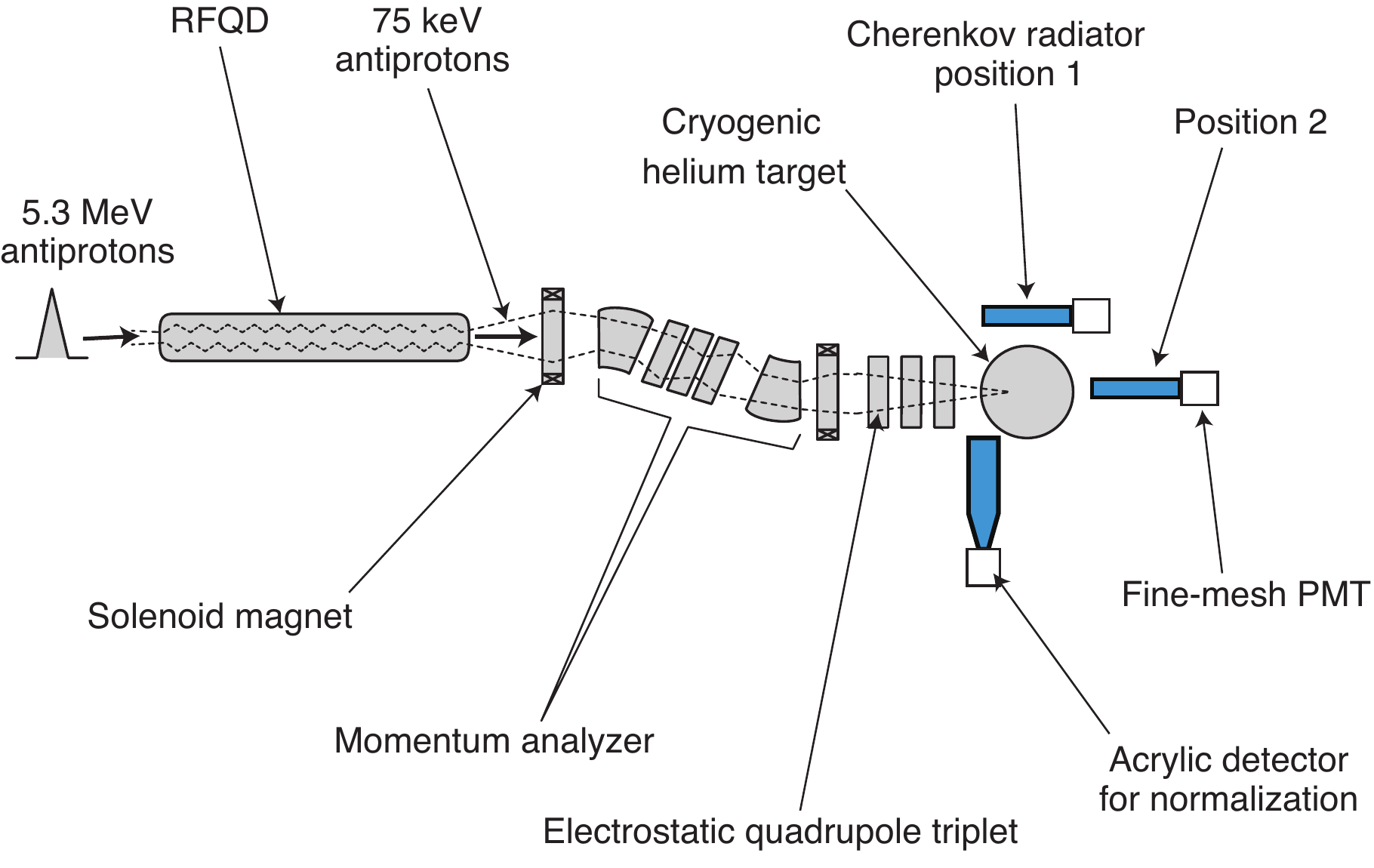}
 \end{center}
 \caption{Experimental layout used to measure the responses of Cherenkov radiators against antiproton annihilations (not to scale). Dashed lines show the trajectory of the antiproton beam. The antiprotons were decelerated to $E=75$ keV by allowing them to pass through a radiofrequency quadrupole decelerator (RFQD). A momentum analyzer beamline selected the 75 keV antiprotons and transported them to a cryogenic target chamber filled with helium gas. The Cherenkov radiators were placed at two locations indicated by positions 1 and 2.
 A separate acrylic Cherenkov radiator read out by a fine-mesh photomultiplier tube was used to measure the intensity of the antiproton beam and normalize the data, see text.}
\label{fig:beamline_crystal_test}
\end{figure}

The schematic layout of the experimental setup used to measure the relative light yields $\Gamma$ of the Cherenkov 
radiators against antiproton annihilations are shown in Fig.~\ref{fig:beamline_crystal_test} \cite{mhori2011,mhori2016}. 
The AD provided a 200 ns long pulsed beam containing $N_{\overline{p}}=(2-3)\times 10^7$ antiprotons with a kinetic 
energy $E=5.3$ MeV and repetition rate $f\sim 0.01$ Hz. The beam was transported along an evacuated
beamline. About 25\% of the beam was decelerated to $E=75$ keV by allowing the antiprotons to pass through a
radiofrequency quadrupole decelerator (RFQD). 
A momentum analyzer beamline selected the $E=75$ keV antiprotons and transported them to a cryogenic target chamber 
filled with helium gas. The antiprotons came to rest in the helium or metallic walls of the target chamber and
underwent nuclear absorption and annihilation.
Past experiments show (see Ref.~\cite{mhori2003-2} and references therein) 
that on average some 1.5 $\pi^+$ mesons, 1.8 $\pi^-$ mesons, and 2 $\pi^0$ mesons 
emerge from a $\overline{p}$+$^4$He annihilation, and that the momentum of most of the pions are distributed 
between $p=0.1$ and 0.9 GeV/c. The $\pi^0$ mesons preferentially decay into two $\gamma$ rays while 
still inside the target. The annihilations also lead to fission of the target nuclei, and the emission of fast 
and slow neutrons. 

\begin{figure}[tb]
\centering
\includegraphics[width=75mm]{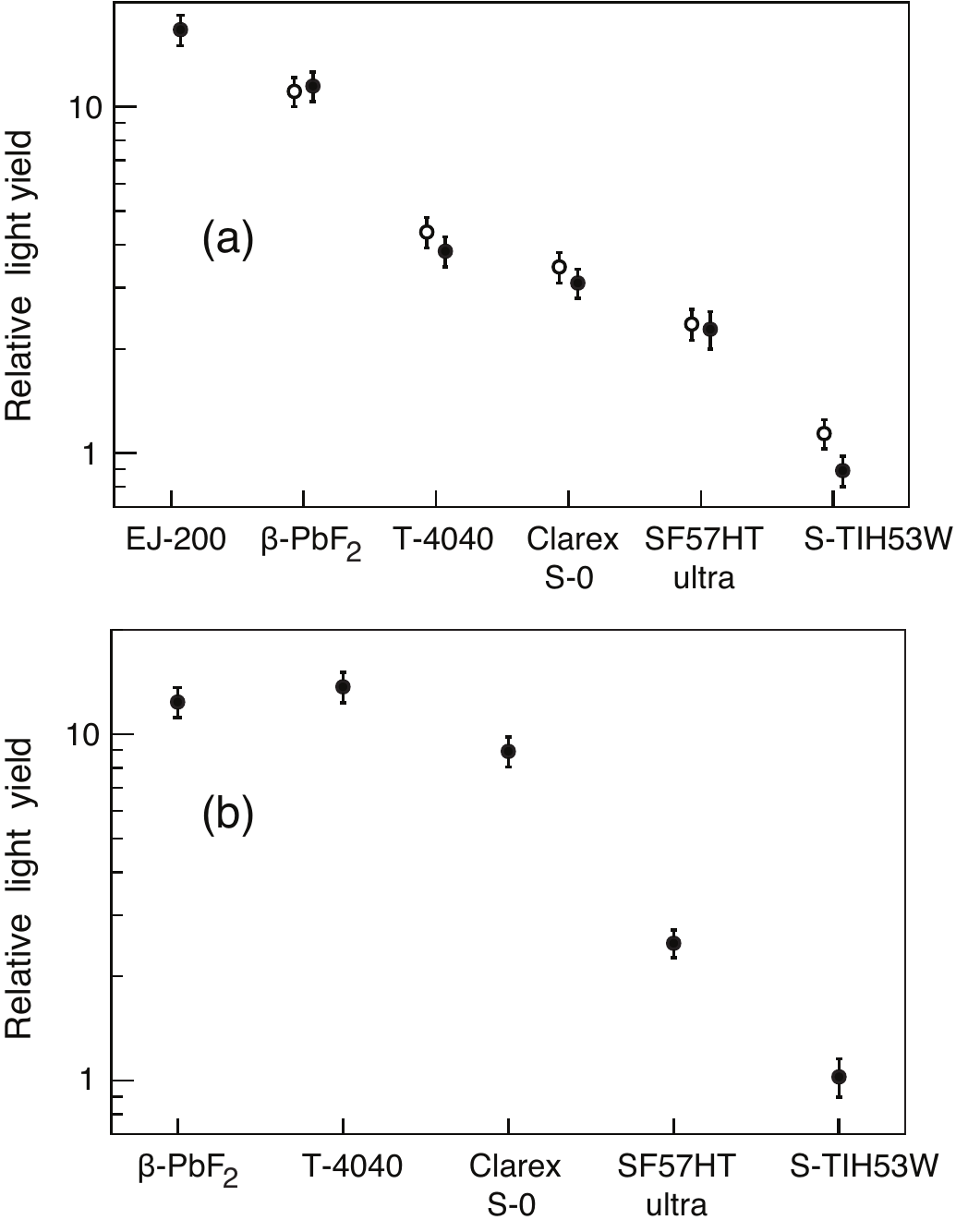}
\caption{Relative light yields of various Cherenkov radiators against antiproton annihilations measured by placing the detector at the two positions 1 (a) and 2 (b) indicated in Fig.~\ref{fig:beamline_crystal_test}. The radiators include lead fluoride ($\beta$-PbF$_2$), 
fused silica (T-4040), UV transparent acrylic (Clarex S-0), lead glass (SF57HTultra) and lead-free high refractive
index glass (S-TIH53W).
For the light yields measured at position 1, no significant differences were seen between detector configurations with (indicated by open circles) and without (filled circles) a silicone light coupler placed between the radiator and a fine-mesh photomultiplier, see text. The light yield of a plastic scintillator (EJ-200) is also shown.}
\label{fig:photon yield}
\end{figure}

A single $\beta$-PbF$_2$ crystal was oriented so that either its short (indicated by position 1 in Fig.~\ref{fig:beamline_crystal_test}) or long (position 2) axis aligned towards the center of the experimental target at a distance $\sim 150$ mm. The Cherenkov light produced in the crystal by the antiproton annihilations were measured by a fine-mesh photomultiplier tube (Hamamatsu Photonics
R5505G-ASSY) \cite{mhori2003-2} with a bialkali photocathode of diameter $d=17.5$ mm and an entrance window made of borosilicate glass. 
The photomultiplier had a quantum efficiency of $>20\%$ at wavelengths $\lambda=380$--440 nm.
The measurements were repeated for the fused silica, acrylic, lead glass, and S-TIH53W glass radiators, and a plastic scintillator 
(Eljen EJ-200) of the same size. The photon yields at position 1 were measured with and without a transparent $t_d=5$ mm 
thick disk made of a two-component silicone rubber (Shinetsu Silicone CAT-103 and KE-103) inserted between the radiator and the 
entrance window of the photomultiplier to improve the optical coupling. The relative intensity of each antiproton pulse was normalized using the readout of a second Cherenkov detector of size 480 mm$\times$160 mm$\times$20 mm made of UV-transparent acrylic (Mitsubishi Rayon, Acrylite000). The detector was positioned 400 mm away from the target, so that the solid angle seen from the target was $\sim 1\%$.

The relative light yields $\Gamma$ of the five radiators and plastic scintillator measured at position 1 are compared in Fig.~\ref{fig:photon yield}(a). Each data point represents the average of measurements collected over 30 antiproton pulses, using three samples of each radiator type. The $\beta$-PbF2 crystals showed the highest light yield that was a factor three larger than the $\Gamma$ values of the fused silica and acrylic radiators. This is due to the short radiation length and high refractive index of $\beta$-PbF2 that allowed the efficient detection of $\gamma$-rays and low-momentum pions. Primarily because of the low transparencies at short wavelengths, the lead glass and S-TIH53W radiators showed light yields that were factors of 5 and 10 smaller compared to $\beta$-PbF$_2$, respectively. No significant differences were seen between the $\Gamma$-values measured with (indicated by open circles) and without (filled circles) the silicone light coupler placed between the radiators and photomultiplier. 

Fig.~\ref{fig:photon yield}(b) shows the light yields measured with the five radiator types placed at position 2. The $\Gamma$-values of the fused silica and $\beta$-PbF$_2$ radiators are now approximately equal, despite the factor $\sim 13$ shorter radiation length of the latter material. This is believed to be due to the lower transparency of $\beta$-PbF$_2$ compared to fused silica which results in the absorption of a significant number of those UV Cherenkov photons that were produced in the part of the crystal closest to the experimental target, that prevent them from reaching the photomultiplier.

\begin{figure}[tbp]
\centering
\includegraphics[width=85mm]{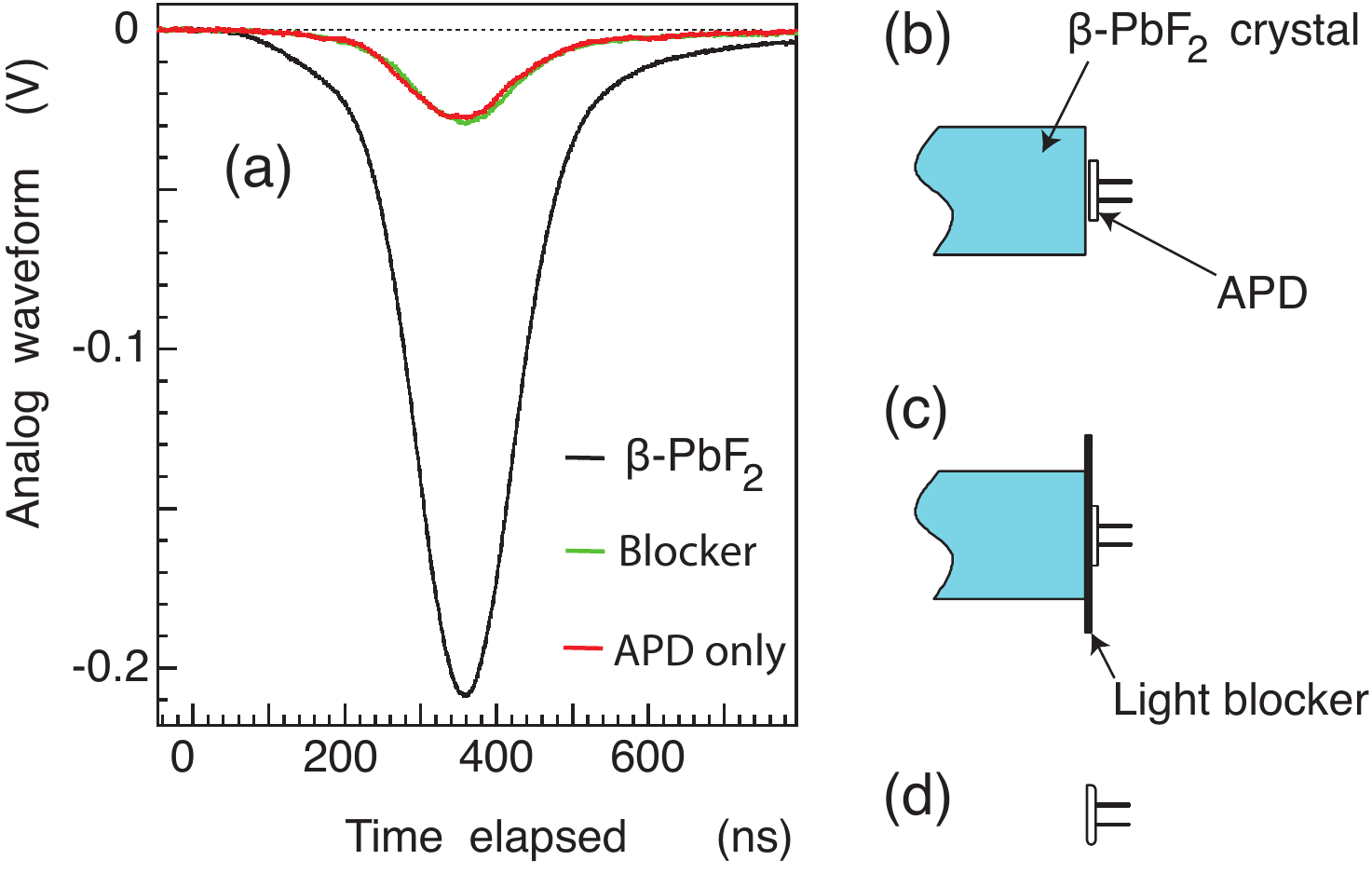}
\caption{(a) Black curve indicates the signal waveform of antiproton annihilations measured by reading out the 
$\beta$-Pb$F_2$ radiators with APD's as shown schematically in (b). It corresponds to the temporal profile of 
the 400 ns long pulsed antiproton beam. The green curve shows the NCE waveform measured by using a rubber 
sheet to prevent the Cherenkov light from reaching the APD as shown in (c). The red curve which is partially 
overlapping with the green one, corresponds to the waveform measured using only the APD without any radiator (d), see text.}
\label{fig:apd_wave}
\end{figure}

\begin{figure*}[tb]
\centering
\includegraphics[width=120mm]{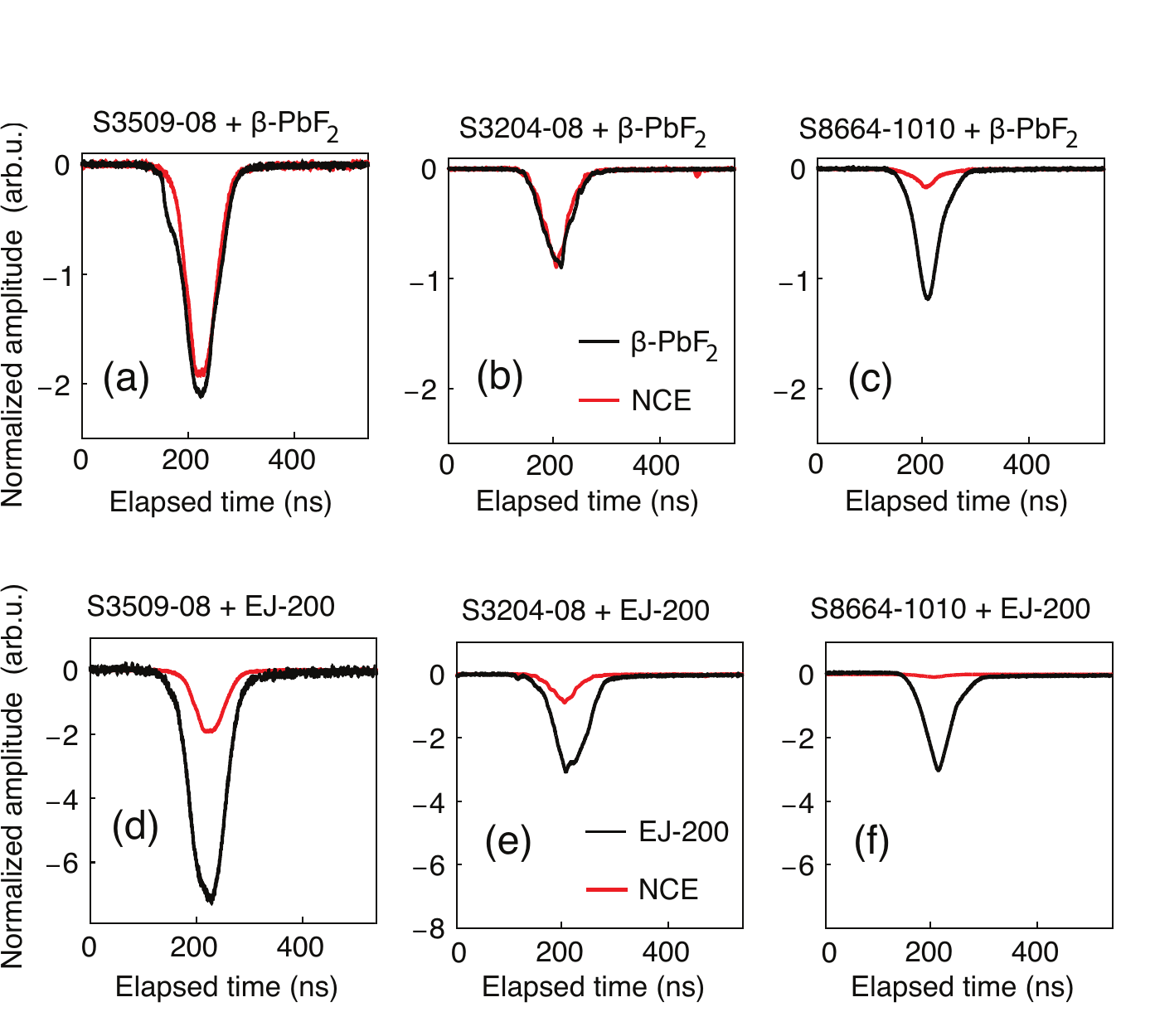}
\caption{Signal waveforms of various detectors measuring antiproton annihilations in the helium target.
Black curves represent the waveforms of the {\it p-i-n} photodiodes (a) S3509-08 and (b) S3204-08, and (c) the APD S8664-1010 coupled to a $\beta$-PbF$_2$ radiator array. The NCE waveforms measured using only the photodiodes without the radiator are plotted using red curves that are superimposed on each spectrum. Signal and NCE waveforms of the light output of a plastic scintillator array read out by (d) S3509-08, (e) S3204-08, and (f) S8664-1010 photodiodes, see text.}
\label{fig:background_photodiode}
\end{figure*}

\subsection{Nuclear counter effects}

We next studied the nuclear counter effect (NCE) caused by secondary particles traversing the readout photodiodes. 
For this the $\beta$-PbF$_2$ array of Fig.~\ref{fig:detector} was positioned at a distance $\sim 550$ mm from the 
cryogenic helium target, with the short axis of the crystals aligned towards the direction where the antiprotons annihilated.
The AD provided a 400 ns long antiproton beam. The black curve of Fig.~\ref{fig:apd_wave}(a) shows 
the analog waveform of the signal obtained by reading out the $\beta$-PbF$_2$ crystals using APD's, and averaging 
the data accumulated over 10 consecutive arrivals of antiproton pulses at the target. A schematic drawing of one of
the five crystal-APD connections in the array is shown in Fig.~\ref{fig:apd_wave}(b). The APD's were biased at a voltage 
$V_b=380$ V. The signal is a superposition of the flash of Cherenkov light in the crystal and the NCE, and shows 
the temporal profile of the beam. To roughly estimate the contribution of the NCE, the measurement was repeated with a 
black rubber sheet placed between the crystals and the APD's to prevent the Cherenkov photons from
reaching the APD's [Fig.~\ref{fig:apd_wave}(c)]. The measured waveform [indicated by the green curve of Fig.~\ref{fig:apd_wave}(a)]
shows that $\sim 15\%$ of the detector signal arises from the NCE. The measurement was repeated with the
$\beta$-PbF$_2$ crystals removed from the detector. No significant change
[red curve of Fig.~\ref{fig:apd_wave}(a)] was seen compared to the measurement with the rubber sheet. 
This indicates that the effect of the crystals shielding the APD's from secondary particle hits was relatively small under the
geometric layout of the experiment.

We next repeated the measurements by using the S3509-08 and S3204-08 {\it p-i-n} photodiodes as well as the 
APD's to read out the five $\beta$-PbF$_2$ crystals. The resulting waveforms are shown using black curves in 
Figs.~\ref{fig:background_photodiode}(a)--(c). The pulse duration of the antiproton beam was reduced to 
$\Delta t\sim 200$ ns during these measurements. Each waveform 
represents the average of data collected from five antiproton pulses. The signal was normalized
against the intensity of the antiproton beam which was measured using a separate acrylic Cherenkov detector 
read out by a fine-mesh photomultiplier tube. The NCE waveforms of each photodiode type measured by 
removing the $\beta$-PbF$_2$ crystals are shown superimposed using red curves.
The fact that the amplitudes of the {\it p-i-n} photodiode waveforms measured with and without the crystals are similar 
shows that most of the detector signal arises from the NCE. This agrees with the results of Monte-Carlo simulations \cite{geant4-1} 
that indicate that each pion hit on a {\it p-i-n} photodiode predominantly produced $>20000$ ion-electron pairs, compared 
to an estimated few hundred photoelectrons collected per hit on the Cherenkov radiator. The amplitude of the 
NCE waveform was proportional to the number of antiproton annihilations. The NCE effectively restricted the
sensitive volume of the detector against charged pion and $\gamma$-ray hits to the $\sim 0.3$--0.5 cm$^3$
depletion regions in the photodiodes, instead of the full 720 cm$^3$ volume of the $\beta$-PbF$_2$ crystal array.
The measurements of the antiproton annihilations were repeated using the EJ-200 plastic scintillator array. 
The signal waveforms of the scintillators read out by the three types of photodiodes are plotted in 
Figs.~\ref{fig:background_photodiode}(d)--(f) using black curves. 
In the case of the {\it p-i-n} photodiode, the NCE contribution [red curves of Figs.~\ref{fig:background_photodiode}(d)--(e)] 
corresponded to $\sim 25\%$ of the total signal.

\begin{figure}[tbp]
\centering
\includegraphics[width=80mm]{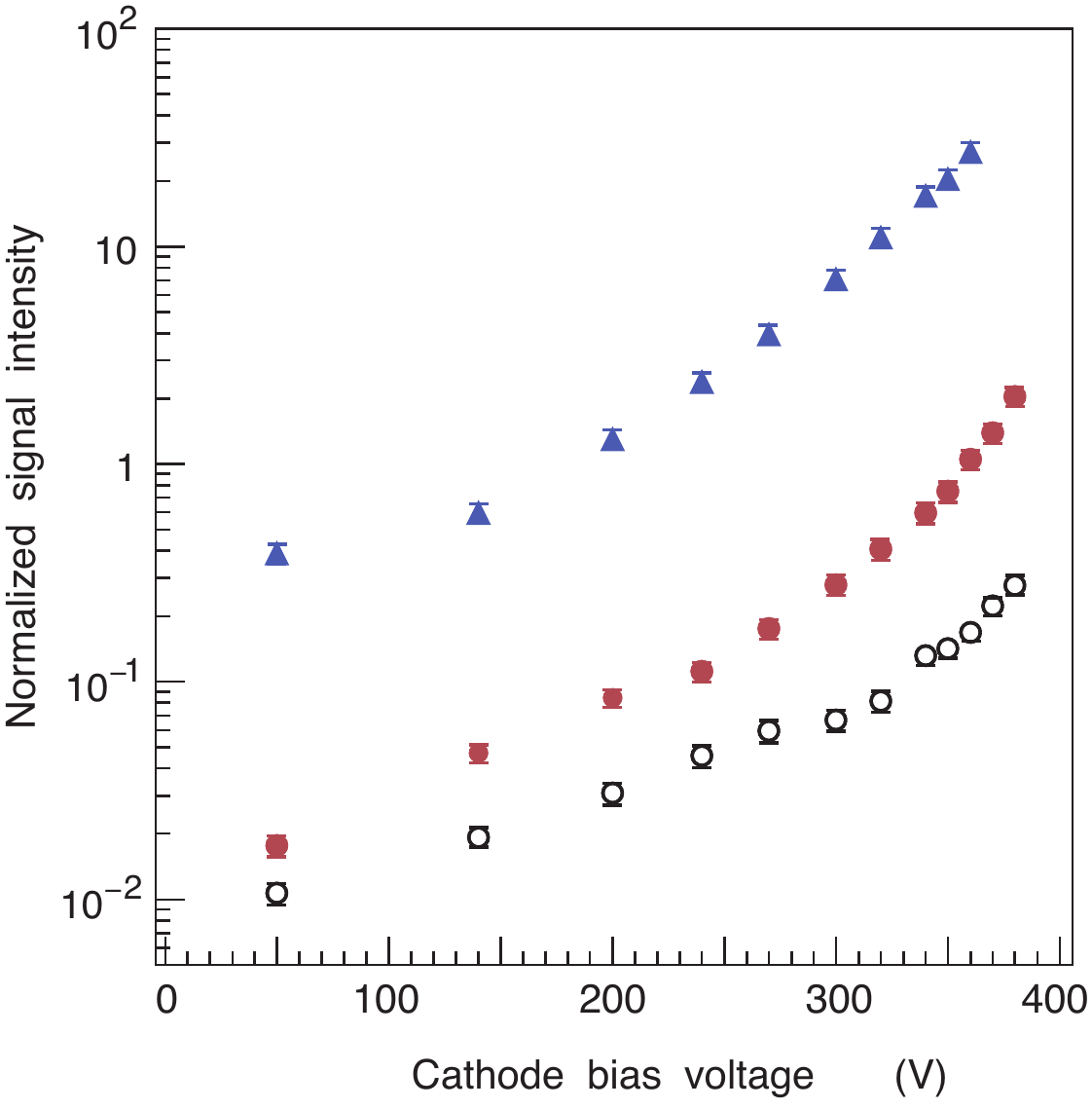}
\caption{Signal intensities of $\beta$-Pb$F_2$ Cherenkov radiators read out by APD's normalized to the intensity of the antiproton beam, as a function of the bias voltage applied to the APD (filled rectangles). Signal intensities of the plastic scintillator array read out by APD's (triangles). 
NCE intensities obtained using only the APD without any radiator (open circles).}
\label{fig:apd_voltage}
\end{figure}

Fig.~\ref{fig:apd_voltage} shows the relative intensities of the APD signals against antiproton annihilations, measured
at bias voltages between $V_b=50$ V and 400 V. Each data point corresponds to the average of data collected from
five antiproton pulses, which were normalized against the intensity of the antiproton beam measured using the
acrylic Cherenkov detector. Measurements were carried out with (indicated by closed circles) and 
without (open circles) a $\beta$-PbF$_2$ radiator crystal coupled to the APD. At low bias voltages $V_b=50$ V which 
corresponded to an APD gain near unity, the NCE constituted $>60\%$ of the total signal. At 
higher bias $V_b=380$ V corresponding to a APD gain $g\ge 80$, the relative intensity of the NCE was 
reduced to $\sim 15\%$ of the Cherenkov detector signal as mentioned above.
A similar tendency of smaller relative NCE at higher gain was seen for APD's reading out plastic scintillators 
(indicated in Fig.~\ref{fig:apd_voltage} by triangles), though some saturation in the signal was seen at $V_b\ge 320$ V. 
At a bias $V_b=380$ V, the NCE corresponded to $\sim 2\%$ of the total signal, as shown in the signal waveforms of Fig.~\ref{fig:background_photodiode}(f). 

\begin{figure}[tb]
\centering
\includegraphics[width=80mm]{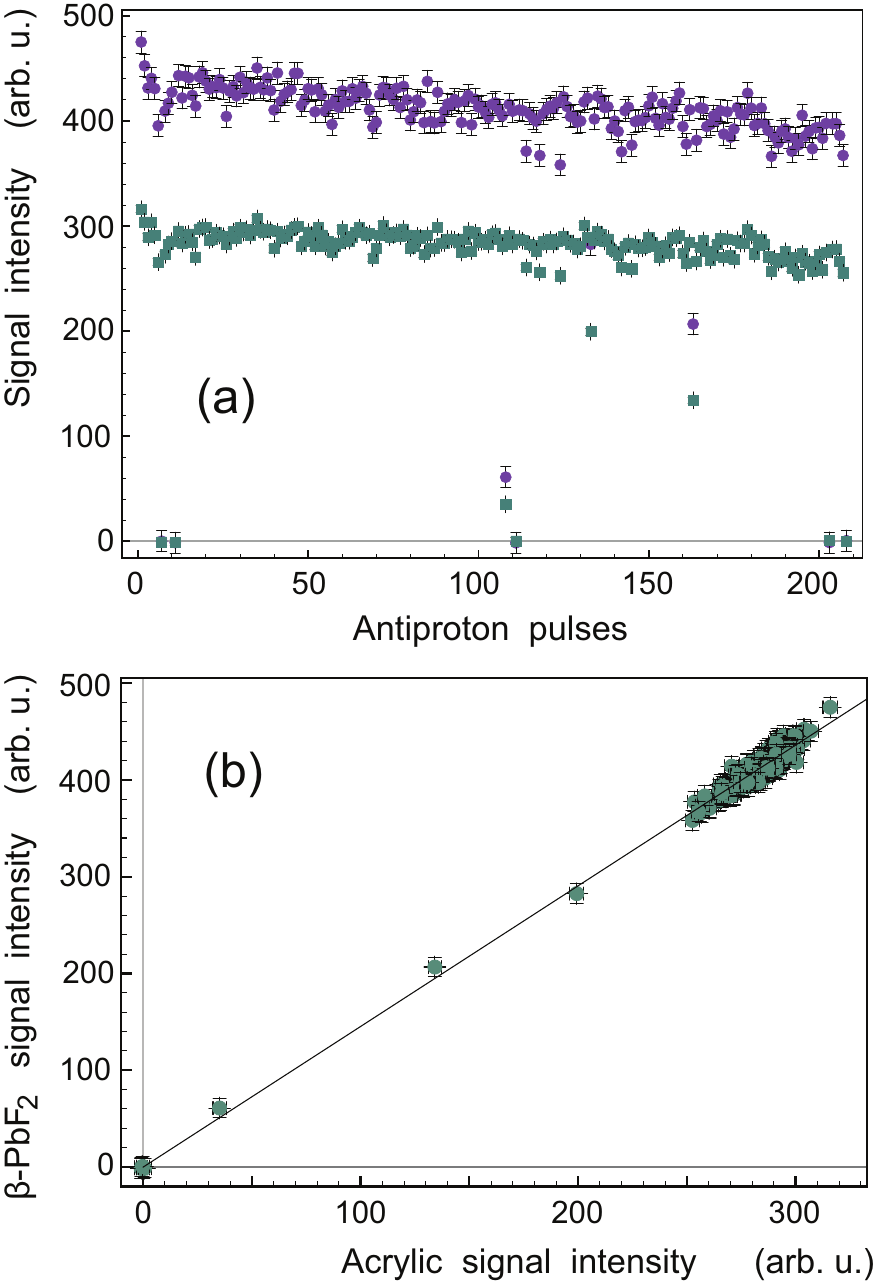}
\caption{(a) Relative intensities of 205 antiproton pulses arriving at the cryogenic helium target,
measured using an array of $\beta$-Pb$F_2$ Cherenkov radiators read out by APD's
(filled circles) and an acrylic radiator read out by a fine-mesh photomultiplier tube
(squares). (b) Correlation between the signal intensities of the two Cherenkov detectors.}
\label{fig:apd_pbars}
\end{figure}

\subsection{Antiproton beam intensity}
\begin{figure}[tb]
\centering
\includegraphics[width=85mm]{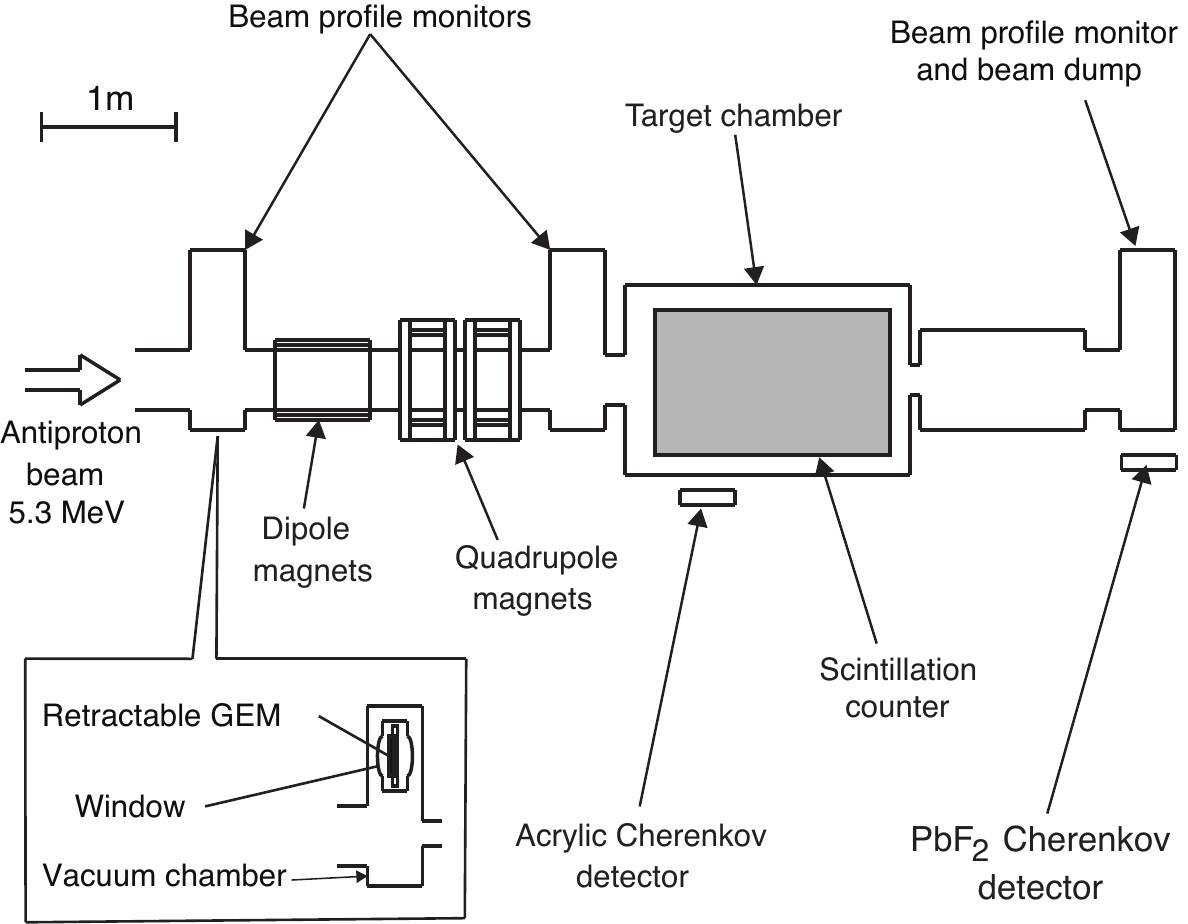}
\caption{Experimental layout to measure the cross section $\sigma^A_{\rm tot}$ of antiprotons with a kinetic energy 
$E=5.3$ MeV annihilating in a carbon target (not to scale) \cite{aghai2018}. A pulsed antiproton beam was allowed to traverse a target chamber that
contained a thin carbon target. The small number
of annihilations that occurred in the target was measured by a scintillation counter. The rest of the antiprotons
struck a profile monitor based on gas electron multipliers (GEMs) located at the end of the experimental setup.
The $\beta$-PbF$_2$ Cherenkov detector was positioned 300 mm away from the GEM detector. The 
short axis of the crystals was aligned towards the direction of the position where the antiprotons annihilated.}
\label{fig:beamline2015}
\end{figure}

Fig.~\ref{fig:apd_pbars}(a) shows the relative intensities of 205 antiproton pulses that were decelerated by the 
RFQD to energy $E=75$ keV over a 8 h period, measured using the $\beta$-PbF$_2$ Cherenkov detector array read out by 
APD's (indicated by filled circles). A bias voltage $V_{b}=380$ V was used. We simultaneously measured the response 
of the acrylic Cherenkov detector read out by a fine-mesh photomultiplier (filled squares), which is known to be highly
linear at these beam intensities \cite{mhori2003-2}. The shot-to-shot fluctuations seen here were primarily caused by
variations in the number of antiprotons that were created in the production target of the AD. 
The correlation between the intensity measurements made by the two detectors [Fig.~\ref{fig:apd_pbars}(b)]
indicates that the linearity of the $\beta$-PbF$_2$ Cherenkov detector array is better than 
$5\%$. The residual nonlinearity is assumed to be mostly due to the contribution of the NCE and
the response of the APD at high light intensities. Time- and temperature-dependent
shifts in the APD gain were estimated to be of order $\sim 2\%$.

Fig.~\ref{fig:beamline2015} shows the layout of an experiment  \cite{aghai2018}
employing the $\beta$-PbF$_2$ Cherenkov detector, in which the total cross section of antiprotons 
with kinetic energy $E=5.3$ MeV annihilating in a carbon target was measured.
In the experiment, radiofrequency cavities were first used to bunch the $(2-4)\times 10^7$ antiprotons circulating
in the AD into six pulses which were distributed equidistantly around its 190 m-circumference.
These pulses of length $\Delta t\sim 50$ ns were then sequentially extracted to the experiment
at intervals of $\sim 2.4$ s, by exciting a kicker magnet located within the AD. The beam was
focused by a quadrupole doublet before entering a 3-m long vacuum chamber containing a
diamond-like carbon target with a thickness of either $t_d=0.7$ or $1.0$ $\mu{\rm m}$. 
Between 1 and 10 annihilations occurred via in-flight nuclear reactions for every $10^6$ antiprotons 
traversing the target foil. The remaining antiprotons emerged from the other side of the foil,
traversed a 2 m long section of evacuated beam pipe, and stuck a profile monitor based on 
gas electron multipliers (GEMs) located at the end of the experiment. This GEM detector
was mounted inside the vacuum pipe of the beamline, and functioned as a beam dump
where all the antiprotons annihilated. The $\beta$-PbF$_2$ Cherenkov detector was positioned 
300 mm away from the GEM detector, with the short axis of the crystals aligned toward the annihilation position 
of the antiprotons. The relative intensity of the antiproton beam measured by the $\beta$-PbF$_2$ detector was
used to appropriately normalize the annihilation counts measured by the scintillation counter to obtain
the cross section. Further details and the results of the experiment are provided in Ref.~\cite{aghai2018}.

\section{Discussions and Conclusions}

In conclusion, we have constructed a Cherenkov detector consisting of an array of five $\beta$-PbF$_2$ crystals of size 30 mm$\times$ 30 mm$\times$160 mm read out by APD's. The detector was used to measure the flux of secondary particles (mostly charged pions and $\gamma$ rays) emerging from the annihilations of high-intensity antiproton beams. When the radiators were positioned so that the majority of the 
secondary particles entered the radiator in the direction parallel to its short axis, the light yield of the $\beta$-PbF$_2$ crystals measured with a bialkali photomultiplier tube was a factor $>2$--3 higher compared to acrylic and glass radiators of the same size. When the secondary particles traversed the long axis of the radiators, however, the light yields of the $\beta$-PbF$_2$ and fused silica radiators became approximately equal.
This is presumably caused by the lower UV transmission of $\beta$-PbF$_2$ compared to fused silica. Cherenkov radiators 
read out using {\it p-i-n} photodiodes showed output signals that mostly consisted of the contribution from the nuclear counter effect 
(NCE) of secondary particles striking the photodiodes. APD's were less affected by the NCE, so that its contribution in the case of
$\beta$-PbF$_2$ readout was $\sim 15\%$ of 
the total signal in the experimental setup used by us. The {\it p-i-n} photodiode and APD readout of plastic scintillators showed that the
NCE contribution was $\sim 25\%$ and $\sim 2\%$ for the two photodiode cases, respectively.
We used the $\beta$-PbF$_2$ Cherenkov radiators read out by APD's in a recent measurement to determine the 
cross section of antiprotons with kinetic energy $E=5.3$ MeV annihilating in carbon targets.

The authors are grateful to the ASACUSA collaboration and the proton synchrotron and Antiproton Decelerator operation teams of CERN. This work was supported by the Max-Planck-Gesellschaft.


\end{document}